\documentclass[10pt,a4paper]{article}
\usepackage[top=2cm,left=7.6cm,footskip=1cm]{geometry}

\usepackage{amsmath,amssymb}

\usepackage{changepage}

\usepackage[utf8x]{inputenc}

\usepackage{textcomp,marvosym}

\usepackage{cite}

\usepackage{nameref,hyperref}

\usepackage[right]{lineno}

\usepackage{microtype}
\DisableLigatures[f]{encoding = *, family = * }

\usepackage[table]{xcolor}

\usepackage{array}

\newcolumntype{+}{!{\vrule width 2pt}}

\newlength\savedwidth

\raggedright
\usepackage[aboveskip=1pt,labelfont=bf,labelsep=period,justification=raggedright,singlelinecheck=off]{caption}

\makeatletter
\renewcommand{\@biblabel}[1]{\quad#1.}
\makeatother

\usepackage{lastpage,fancyhdr,graphicx}
\usepackage{epstopdf}
\usepackage{multirow}
\usepackage{subcaption}

\begin{document}
\vspace*{0.2in}

\begin{flushleft}
{\Large
\textbf\newline{Visualizing spreading phenomena on complex networks}
}
\newline
\\
Christian Schulz\textsuperscript{1*}
\\
\bigskip
\textbf{1} Computational Social Science, ETH Zurich, Zurich, Switzerland
\\
\bigskip

* cschulz@ethz.ch

\end{flushleft}
\section*{Abstract}
Graph drawings are useful tools for exploring the structure and dynamics of data that can be represented by pair-wise relationships among a set of objects. Typical real-world social, biological or technological networks exhibit high complexity resulting from a large number and broad heterogeneity of objects and relationships. Thus, mapping these networks into a low-dimensional space to visualize the dynamics of network-driven processes is a challenging task. Often we want to analyze how a single node is influenced by or is influencing its local network as the source of a spreading process. Here I present a network layout algorithm for graphs with millions of nodes that visualizes spreading phenomena from the perspective of a single node. The algorithm consists of three stages to allow for an interactive graph exploration: First, a global solution for the network layout is found in spherical space that minimizes distance errors between all nodes. Second, a focal node is interactively selected, and distances to this node are further optimized. Third, node coordinates are mapped to a circular representation and drawn with additional features to represent the network-driven phenomenon. The effectiveness and scalability of this method are shown for a large collaboration network of scientists, where we are interested in the citation dynamics around a focal author.
\section*{Introduction}

With an ever-growing availability of large networked data \cite{lazer2009life,john2014big} and the need for a better understanding of complex network-driven phenomena \cite{barabasi2012network,pastor2015epidemic}, graph visualization techniques are important tools for a wide range of scientific disciplines that analyze social, biological or technological networks. Drawing networks helps us to explore the complex relationships between individual graph nodes and visually support quantitative findings.

 Classical node-link diagrams are increasingly difficult to draw with a rising number of nodes and edges connecting the nodes. Large real-world networks often exhibit small-world properties, i.e., a low average shortest path length and a high clustering coefficient \cite{watts1998collective}. For this type of networks, it is not feasible to find a low-dimensional embedding for an accurate drawing. Networks with a high clustering can potentially be visualized in a meaningful way so that the graph drawing reveals a rough community structure, where sets of nodes that are plotted close together are interpreted as a cluster which is to some extent separated from the rest of the network. However, when networks exhibit a low average shortest path length, i.e., when nodes reach other nodes using only a few links, graph drawings may end up with a single cluster with an unclear link structure. Specifically, for the study of processes where a node influences its neighboring nodes, for example spreading phenomena, these small-world links may play a crucial role.

 Consequently, a visualization method is proposed that focuses on one node at a time and preserves network distances to all other nodes. Since this is a less demanding task than preserving network distances between all pairs of nodes in a low-dimensional space, it is thus applicable to extensive real-world networks (more than 100 million edges).

 Deciding for a specific focal node should still be feasible interactively so that it becomes necessary first to compute an approximate global solution. Here, we can build on many established network layout algorithms \cite{hu2015visualizing}: Popular force-based approaches determine optimal node positions by an iterative method that finds a balance between attractive forces (towards neighboring nodes) and repulsive forces (away from any other node). Examples of such algorithms are Fruchtermann-Reingold \cite{fruchterman1991graph}, Hu \cite{hu2005efficient} (efficient repulsion force approximation), ForceAtlas2 \cite{jacomy2014forceatlas2} (different force model, adaptive cooling scheme) or OpenORD \cite{martin2011openord} (multi-level layout). On the other hand, distance-based methods directly minimize the error between network distance and distance in the projected space. Examples of this class of algorithms are Multidimensional scaling (MDS) \cite{torgerson1952multidimensional} or Pivot MDS \cite{brandes2006eigensolver} (efficient MDS approximation). All of these algorithms perform computations in Euclidean space. In contrast, \cite{boguna2010sustaining} maps a scale-free network \cite{barabasi1999emergence} to hyperbolic space, where the radial coordinate corresponds to an exponentially decreasing node degree and together with the angular coordinate approximates network distance. Similarly, \cite{munzner1997h3} utilized a hyperbolic representation to draw a focused graph hierarchy while still preserving a global context. Comparing different spaces, \cite{du2017isphere} found that a spherical mapping is evaluated visually favorably to a planar or hyperbolic projection. In this work, we first find optimal coordinates in spherical space and then, after choosing a focal node, project to a polar coordinate system with the focal node in the center.

 By focusing on the potential source of a spreading process, we would expect to see a diffusion with increasing distance to the focal node. For example, \cite{brockmann2013hidden} combined data of a global spread of infectious diseases with the world-wide airport traffic network and achieved a much better prediction of the contagion process than what could be estimated from geographic distance. Visually, when putting the outbreak location in the center of a focal network layout where radius corresponds to network distance, they could observe a circular wavefront of reported cases propagating through the network. In contrast to epidemic spreading, other contagion processes such as social influence seem to be more difficult to measure. In the social sciences, studies would typically treat subjects as independent, since (densely connected) social network data was not available. However, any interaction with friends, colleagues, neighbors etc. may influence their behavior. For example, \cite{christakis2007spread} measure the chance of becoming obese, if a friend became obese. An accompanying social network drawing \cite{christakis2009social} illustrates growing clusters of obesity. In this work, our main example instead concerns the diffusion of knowledge (alternatively called or related to the spread of information, flow of ideas or adoption of innovations). We look at the social network of scientists and visually explore its relationship to the spread of scientific ideas \cite{crane1972invisible}. Since typical network drawings of science mostly concentrate on revealing community structure \cite{boyack2005mapping,rosvall2008maps,borner2012design}, we are more interested in whether we can make any network-driven diffusion process visible. To do so, we examine the spread of citations on the co-author network, i.e., the first time when an author cites another author's work. 

\begin{figure}
\begin{adjustwidth}{0.0cm}{-0mm}
\includegraphics[width=12.0cm]{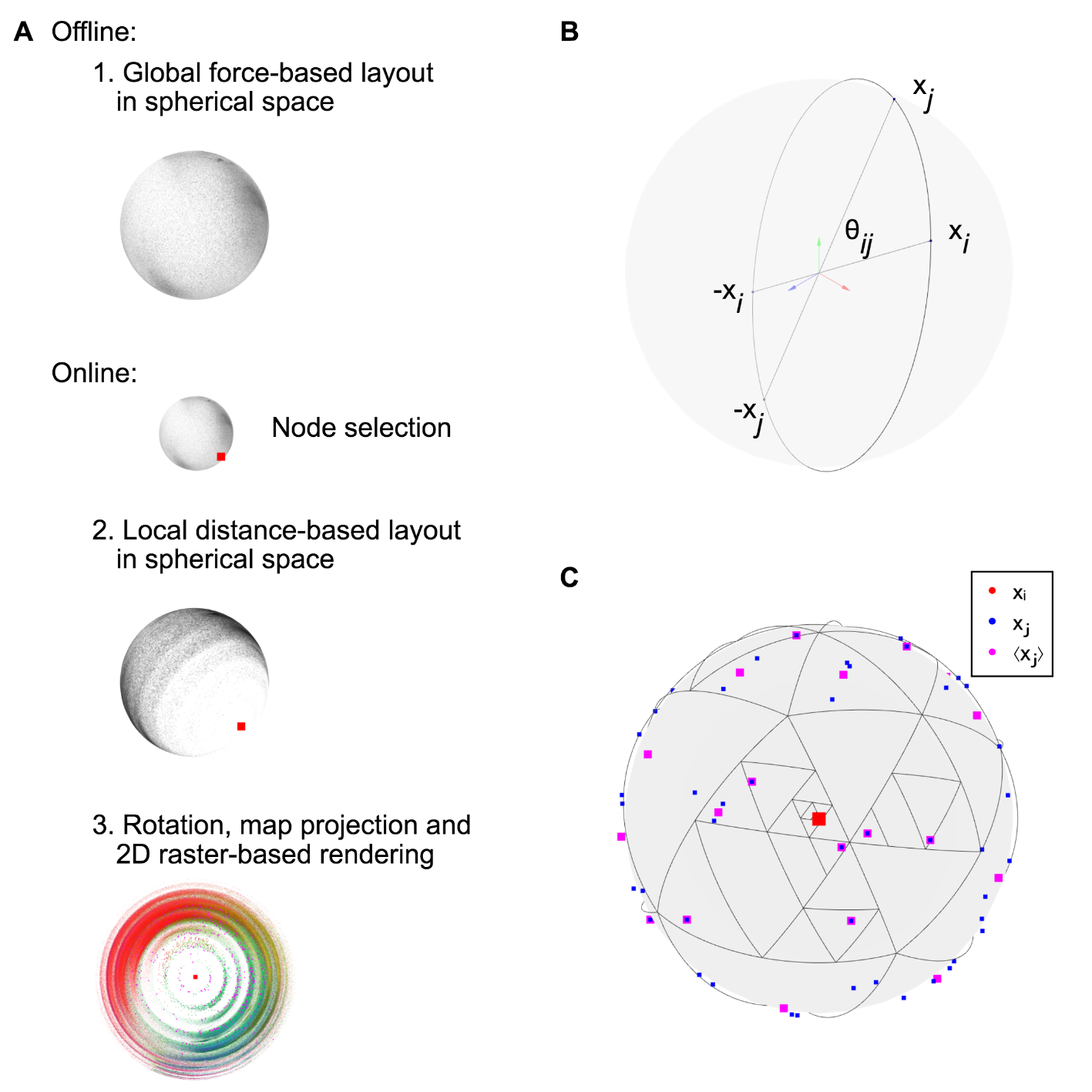}
\end{adjustwidth}
\caption[Schematic overview and force computation in spherical space]{\textbf{Schematic overview and force computation in spherical space.} \textbf{(A)} The network visualization consists of three main steps: 1. For all nodes, a force-based graph layout algorithm computes initial 3D coordinates on the unit sphere. Then, any node of the graph may be chosen from which a focal network visualization shall be generated. 2. Node coordinates are further optimized towards a better network distance preservation to the focal node. 3. The sphere is rotated to center the focal node, projected to 2D and rendered with additional node information of interest. \textbf{(B)} The distance between two node coordinates $\hat x_i$ and $\hat x_j$ on the sphere is simply the angle between them in radians. Attraction towards a node is computed by interpolating on the arc that connects a pair of nodes. Repulsion of $\hat x_i$ away from $\hat x_j$ can be solved by additionally using $- \hat x_j$ or $- \hat x_i$ for the interpolation. \textbf{(C)} For an efficient repulsion force computation, a sub-divided regular icosahedron is used to approximate repulsion forces on $\hat x_i$ from any other node $\hat x_j$ by combining them into a center $\langle x_j \rangle$ for each triangle cell and sub-division level.}
\label{focalLOverview}
\end{figure}

\section*{Methodology}

\subsection*{System overview}

The objective of a network layout algorithm is to find a two-dimensional projection of the graph where the Euclidean distance between any node pair corresponds to the network distance, i.e., the shortest path between the two nodes. Since this is usually only possible for very simple graphs, and our goal is to explore networks that potentially consist of millions or billions of nodes and edges, we need to design a method that creates visual insights even with an approximate solution and is computationally feasible.

 Therefore, we concentrate the visualization on one node at a time, by drawing a focal network layout that shows the network structure from the perspective of a single node. Computational constraints mainly drive the second important design choice: an algorithm that includes the processing of gigabytes of information is difficult to realize as an interactive exploration. Thus, there needs to be a pre-processing phase, which later allows rendering layouts with a different focus quickly. The proposed method divides this task into three main steps, as depicted in Fig.~\ref{focalLOverview} A.

 First, we apply a layout algorithm that tries to find optimal coordinates for all nodes. Since this may lead to only a very approximate solution and potentially contains large distance errors for individual nodes, a second step further optimizes distances to the node of interest. When another node is selected, computed coordinates of the first step can be reused as a starting point. The third step deals with the problem of visualizing millions of nodes, together with additional information. 

\subsection*{Spherical space}

 Node coordinates are typically computed in two-dimensional Euclidean space since the final drawing is mostly in two dimensions, too. In this work, coordinates are represented in spherical space. This has the advantage that a network layout does not necessarily have a single center where there will be nodes at the "border" after which there is only empty space. Focusing on such "border" nodes would mean that all other nodes are distributed in one dominant direction and other directions are left empty. We could circumvent this by wrapping 2D Euclidean space as a torus. However, using a sphere and its projection to 2D, centered at the focal node, seems to be more intuitive. 

 Positions of nodes are represented as 3D vectors with a length constraint of 1 and thereby reside on the surface of a unit sphere centered at the coordinate origin (see also Fig.~\ref{focalLOverview} B):

\begin{equation}
x \in R^3,\; \left\| x \right\|=1
\end{equation}
Normalized vectors are denoted as $\hat{x} = x/||x||.$ The spherical distance between two nodes can be computed as

\begin{equation}
\theta_{ij} = \arccos (\hat x_i \cdot \hat x_j),\;\; 0 \le \theta \le \pi.
\end{equation}
Moving to another position on the sphere surface can be realized by a spherical interpolation between two position vectors \cite{shoemake1985animating}:

\begin{equation}
\operatorname{slerp}(\hat x_i,\hat x_j,\alpha)= \frac{\sin ((1-\alpha) \theta_{ij})}{\sin \theta_{ij}} \hat x_i + \frac{\sin (\alpha \theta_{ij})}{\sin \theta_{ij}} \hat x_j,\;\; 0 \le \alpha \le 1
\end{equation}
In contrast to \cite{kobourov2005non}, where a tangent space is created, our method is free of distortion and computationally efficient.

\subsection*{Global network layout}

The force-based Fruchtermann-Reingold \cite{fruchterman1991graph} layout algorithm is used to compute the initial node coordinates. In principle, any network layout algorithm may be chosen at this step. However, the selected approach is simple to implement, adaptable to spherical space and a good starting point for adding methods that improve computational efficiency or quality of the layout. In general, a force-based algorithm is an iterative method, where at each step, node positions are updated by applying attractive and repulsive forces to pairs of nodes. Attractive forces are used to minimize the distance between nodes that are connected by an edge. We also need repulsive forces that push a node away from any other node; otherwise, all nodes would end up in the same coordinates.

 For a node $i$, we can compute a new position based on attraction to its direct network neighboorhood $N(i)$ by 

\begin{equation}
x_a(i) = \sum^{N(i)}_j \theta_{ij}^2 \operatorname{slerp}(x_i, x_j, \min(1, \theta_{\max} / \theta_{ij} )), 
\end{equation}
where $\theta_{\max}$ specifies the maximum step length. Similarly, repulsion caused by node $j$ is defined as:

\begin{equation}
x_r(i, j)=
 \begin{cases}
   \operatorname{slerp}(x_i, -x_j, \theta_{\max} / (\pi - \theta_{ij})), & \text{if $\pi - \theta_{ij} \leq \theta_{\max}$ };\\
   \operatorname{slerp}(-x_j, -x_i, (\theta_{\max} + \theta_{ij} - \pi) / \theta_{ij}), & \text{otherwise}.
 \end{cases}
\end{equation}
For a node $i$ the overall new position based only on repulsion is then

\begin{equation}
x_r(i) = \sum_{j \neq i} \frac{1}{\theta_{ij}} x_r(i, j).
\end{equation}
The overall new coordinates are the mean of both normalized position vectors:

\begin{equation}
\hat{x}_{i} = \frac{1}{2} (\hat{x}_a(i) + \hat{x}_r(i)) 
\end{equation}
Vector addition is possible since all individual vectors are maximally $\theta_{\max}$ away, which can be specified such that the curvature of sphere can be assumed to be approximately planar. A difference to Fruchtermann-Reingold is that attraction and repulsion are equally weighted, which supports the balance between both forces such that the nodes are spread out over the whole sphere surface. Consequently, a parameter which defines the optimal edge length (they refer to it as $k$) is not needed and is implicitly determined due to the finite space the sphere surface provides for the nodes. Parameter $\theta_{\max}$ can be adapted at each simulation step. Here, a very simple slow-down is implemented, which linearly decreases to 0 over the whole simulation time $t$:

\begin{equation}
\theta_{\max}(t) = (1 - t) \theta_{\max}(0),\; 0 \le t \le 1.
\end{equation}
Before the simulation starts, positions are initialized randomly, with $x=(X_1, X_2, X_3)$, where $X$ is a normally distributed random variable. \cite{muller1959note} shows that the resulting (normalized) points are uniformly distributed on the surface of the sphere.

Repulsion forces need to be evaluated between any pair nodes, which is computationally too expensive for a high number of nodes. Instead, \cite{barnes1986hierarchical} offer an approximation used in n-body simulations, which is based on a quadtree and reduces the computational complexity from $O(|V|^2)$ to $O(|V|log |V|)$, where $|V|$ is the number of nodes.

 In spherical space, we can use a regular icosahedron with 12 vertices and 20 equilateral triangular faces (Fig.~\ref{focalLOverview} C). Each triangle can then be subdivided into 4 likewise equilateral triangles by creating 3 new vertices that bisect the 3 triangle edges. The icosahedron is constructed by two pole vertices at $(0,0,\pm 1)$ and ten more vertices that form two pentagons: 

\begin{equation}
\cos(arctan{\frac{1}{2}}) \bigg(\cos(i \frac{\pi}{5}),\; \sin(i \frac{\pi}{5}),\; \frac{(-1)^i}{2}\bigg),\; i=0,1,...,9.
\end{equation}
Before the force computation starts, all nodes are added to each subdivision level. At the root level, we determine which of the 20 triangles is intersected by the ray formed by (0,0,0) and node position $\hat x$. Not all triangles need to be tested; a transformation of $\hat x$ into an angle $\operatorname{atan2}(\hat x_2, \hat x_1)$ and a $\hat x_3$-coordinate reduces the maximum number of tests to 2. The intersection point that lies in the triangle formed by vertices $v_1$, $v_2$, $v_3$ and the triangle plane normal vector $n_t$ is computed by

\begin{equation}
\dot x = \frac{v_1 \cdot n_t}{\hat x \cdot n_t} \hat x.
\end{equation}
Then, starting with the second tree level, barycentric coordinates are used to describe $\dot x$ relative to one of the four sub-triangles, such that 

\begin{equation}
\dot x= \lambda_1 v_1 + \lambda_2 v_2 + \lambda_3 v_3,\; \lambda_1+ \lambda_2 + v_3 = 1.
\end{equation}
Using the property that barycentric coordinates are equivalent to the ratio of areas of the three triangles spanned by $v1$, $v2$, $v3$ and $\dot x$, we solve

\begin{equation}
\lambda_1 = \frac{||(v_2 - \dot x) \times (v_3 - \dot x)||}{||(v_3 - v_2) \times (v_3 - v_1)||},\; \lambda_2 = \frac{||(v_3 - \dot x) \times (v_1 - \dot x)||}{||(v_3 - v_2) \times (v_3 - v_1)||},\; \lambda_3=1-\lambda_1-\lambda_2.
\end{equation}
On each subdivision level, one of the four containing triangles has to be selected. In barycentric coordinates, the decision borders lie at $\lambda$ = 0.5. Barycentric coordinates of $\dot x$ for the containing triangle can be found by simple linear interpolation of the parent triangle. At each level, the mass center of the triangle is computed by adding $\dot x$. At the leaf level, a list of individual vertices is stored.

 When computing the repulsion target position for for a node, we traverse the triangle tree, starting with the twenty root triangles, and compute the spherical distance of the triangle's mass center (if it contains any nodes) to the position of the node. If this distance exceeds a specified threshold $\theta_{\operatorname{quad}}$ , we can use the mass center to compute the repulsion, otherwise we continue to traverse its children. When arriving at a leaf, repulsion has to be computed for each vertex individually. 

A large number of nodes and edges may make it difficult for the simulation to not get stuck in a local minimum. The idea of a multi-level network layout is to start with a coarser version of the graph, containing only a few nodes, for which it is easier to find an optimal layout, and then, at each layout step, add more nodes until the original network structure is restored. That means we need to find an order of all nodes in which they are added back to the graph. An edge is added as soon as the two connected nodes are included. For temporal graphs, we could use the node creation timestamp to find an order of nodes. Nodes that have the same timestamp are added in a randomized order. For other networks, a random walk on the network is applied by randomly choosing a neighbor node and with a certain probability to jump to another random node. In case the network consists only of one component, the random node is chosen among the already visited nodes to prevent that the layout is performed on disconnected network components.

\subsection*{Focal network layout}

The focal network simply corrects the distance errors between any node $\hat x_i$ and the focal node $\hat x_f$. Therefore, we compare the spherical distance $\theta_{if}$ resulting from the global network layout with the actual network distance $d_{if}$. A shortest path computation is performed to determine all $d_{if}$. For an unweighted network, it is a breadth-first search starting at the focal node, with a computational complexity of $\Omega(|V| + |E|)$. We can then move each node towards its ideal distance to the focal node by

\begin{equation}
x_s(i)=
\begin{cases}
  \operatorname{slerp}(\hat x_i, \hat x_f, \alpha \frac{\theta_{if} - min(1, d_{if} / d_{\max}) \pi}{\theta_{if}}), & \text{if $\theta_{if} \geq \frac{d_{if}}{d_{\max}} \pi$};\\ 
  \operatorname{slerp}(\hat x_i, - \hat x_f, \alpha \frac{min(1, d_{if} / d_{\max}) \pi - \theta_{if}}{\pi - \theta_{if}}), & \text{otherwise}.
\end{cases}
\end{equation}
where $\alpha$ is the interpolation parameter between 0 (no node movement) and 1 (all nodes at the same network distance form a ring). $d_{\max}$ specifies the network distance that corresponds to the maximum spherical distance ($\pi$) and is determined by solving the linear system of all distances $d = \frac{d_{\max}}{\pi} \theta$ of any node pair using least squares.

\subsection*{Projection and rendering}

We perform a 3D rotation such that the focal node $\hat x_f$ moves to the projection center $\hat x_p=(0,0,1)$ in a right-handed coordinate system where we project on the $x_1 x_2$-plane and look towards the negative $x_3$-coordinate. For any node, a new position is found by applying Rodrigues' rotation formula: 

\begin{equation}
\dot{x} = (\hat x_f \cdot \hat x_p) \hat x + \sqrt{1-(\hat x_f \cdot \hat x_p)^2} (\hat{e} \times \hat x) + (1 - \hat x_f \cdot \hat x_p) (\hat e \cdot \hat x) \hat e,\;\; e = \hat x_f \times \hat x_p.
\end{equation}
Projecting a sphere surface on a two-dimensional space is a long-standing topic for creating maps of the earth surface. Here, we choose a Lambert azimuthal equal-area projection that distorts angles, but preserves distances to the projection center (the focal node) and areas to be able to compare different parts of the layout in size and density. A position $\hat x$ is projected to two dimensions by:

\begin{equation}
(\dot x_1, \dot x_2) = (z \hat x_1, z \hat x_2),\;\; z = \frac{1}{2} \sqrt{\frac{2}{1+\hat x_3}},\;\; \dot x_1, \dot x_2 \in [-1, 1].
\end{equation}
Instead of drawing a high number of individual nodes or even edges, we create a 2D histogram that counts nodes at a projected 2D position. Similar to \cite{cottam2013abstract}, this allows for a more controlled color shading of network areas of different densities and is computationally much more efficient. A histogram bin corresponds to a pixel of the raster image on which the network is rendered. Node positions may cover multiple bins for better visibility. A pixel's transparency value is computed as $\frac{1}{1+n}$, where $n$ is the number of nodes in the bin. The output is therefore independent of the number of nodes.

 For 3D renderings of large networks, a 2D texture image for the sphere surface may be created. Projecting the node positions on the texture is also achieved by Lambert azimuthal equal-area mapping. To avoid distortions that are mostly visible at the coordinate opposite of the projection center (where the outer ring of the projection would meet in a single point), the projection is split into two half spheres with two opposing projection centers.   

\begin{table}
\caption[Network data used in experiments]{\textbf{Network data used in experiments. } "ws(...)" are random graphs generated by a Watts-Strogatz model with parameters: number of nodes, node degree and rewiring probability. "dw256A" and "qh882" are from the "University of Florida sparse matrix collection" \cite{davis2011university}. For comparison, layouts of these graphs resulting from other algorithms can be found in \cite{hu2015visualizing}. "WoS" networks are based on Clarivate Analytics' Web of Science \cite{webofscience}, covering publication records from a wide range of scientific disciplines from 1950 to 2015. Co-authorship data is generated by a name disambiguation algorithm presented in \cite{schulz2014exploiting}. Additionally, the same method is applied to extract a co-authorship network from the Microsoft Academic Graph \cite{sinha2015overview}. WoS co-authorships accumulated until the year 2010, MAG co-authorships until 2016. Execution times, layout quality measures $\langle \hat \theta \rangle$ (normalized average edge length, lower is better) and $\rho(d,\theta)$ (distances correlation, higher is better) are the mean and standard deviation of 50 simulation runs. Execution times were measured on 2x12 core machines with 47 concurrent threads. Networks with $\le$ 1,000 nodes were layouted in 500 simulation steps, while larger network layouts stopped after 250 steps.}
\label{focaldatasourcestable}
\resizebox{\textwidth}{!}{%
\begin{tabular}{ lrrrrr }
Name & \# nodes & \# edges & $\langle \hat \theta \rangle$ & $\rho(d,\theta)$ & time in secs \\
\hline
Grid 10x10 & 100 & 180 & .18±.01 & .90±.05 & 5.86±.23 \\
ws(15, 4, .1) & 15 & 30 & .45±.00 & .92±.01 & 2.10±.19 \\
ws(1000, 4, .02) & 1,000 & 2,000 & .06±.00 & .70±.02 & 5.63±.12 \\
dw256A & 512 & 1,004 & .12±.01 & .64±.08 & 5.60±.12 \\
qh882 & 882 & 1,533 & .07±.00 & .70±.02 & 5.63±.12 \\
WoS Sociology co-authorships & 1,850 & 3,063 & .09±.00 & .64±.02 & 3.15±.09 \\
WoS Math co-authorships & 25,356 & 56,111 & .11±.01 & .58±.01 & 14.49±.33 \\
Wos Biology co-authorships & 377,029 & 2,589,377 & .22±.01 & .62±.01 & 214.88±1.27 \\
WoS co-authorships	& 2,490,473 & 28,171,593 & .11±.01 & .69±.01 & 1,934.69±36.85  \\
MAG co-authorships	& 6,046,078 & 63,106,848 & .12±.01 & .69±.01 & 4,113.16±400.63 \\
WoS citations & 38,919,796 & 728,348,927 & .13±.01 & .73±.01 & 29,858.99±1,180.67 \\
\hline
\end{tabular}}
\end{table}

\section*{Results}

Experiments were conducted with network data summarized in Table \ref{focaldatasourcestable}. Small, artificial networks were selected to analyze different aspects of the algorithm, while results for large social networks were provided to show the effectiveness of such an approach. We start with a qualitative evaluation that gives an overview of potential visual insights that can be achieved, followed by a quantitative evaluation concerning error minimization, sensitivity to local minima and performance measurements of the implementation.

\begin{figure}
\begin{adjustwidth}{-6.0cm}{-0mm}
\includegraphics[width=18.0cm]{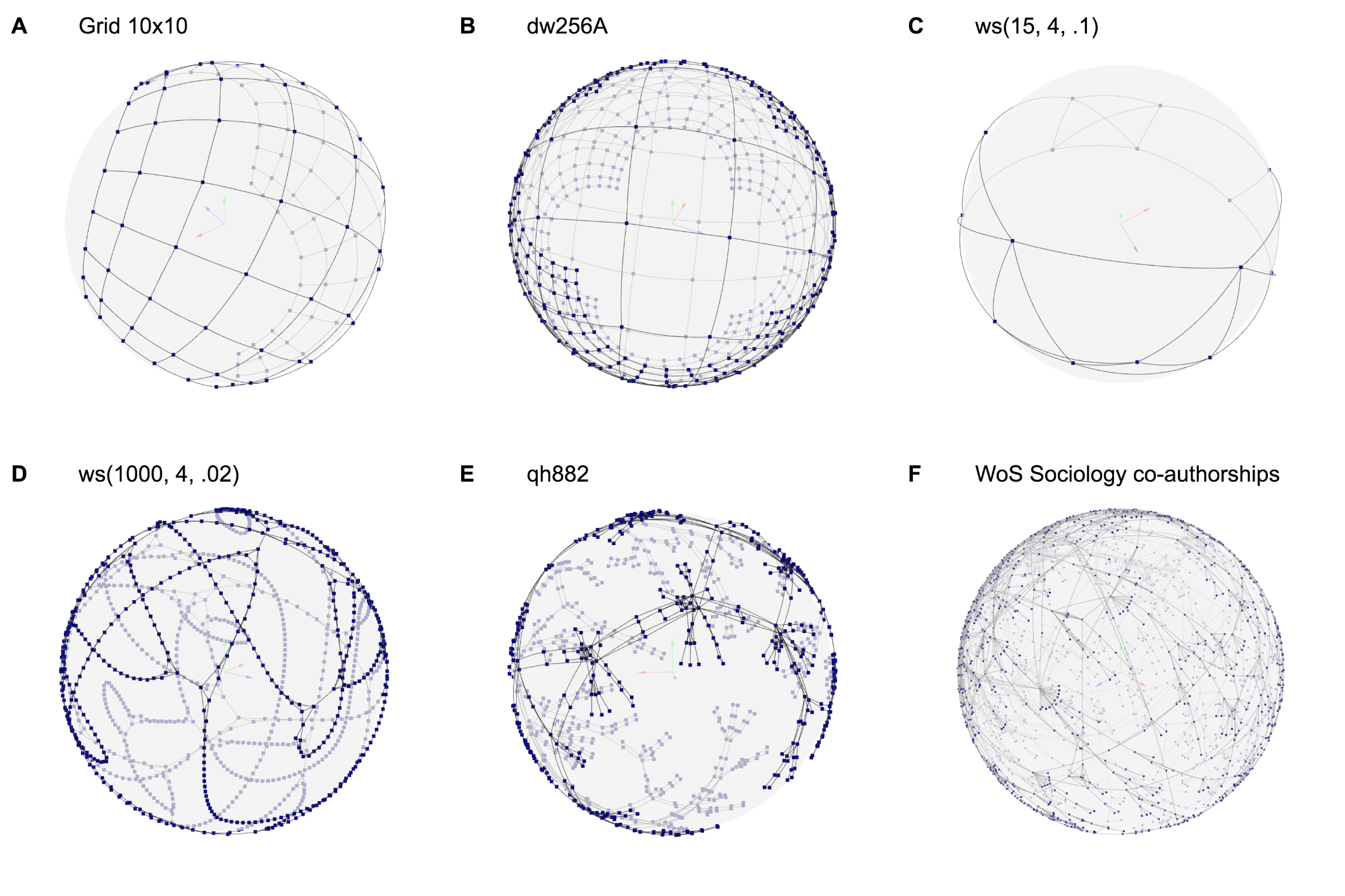}
\end{adjustwidth}
\caption[Spherical layout of small networks]{\textbf{Spherical layout of small networks.}  Network data as described in Table \ref{focaldatasourcestable}. Networks are rendered in 3D space, edges are drawn as great-circle distances. 3D perspective is chosen to taste. An interactive version of the plot allows the user to rotate and zoom the sphere. \textbf{(A)} A grid layout represents a Manhattan distance metric and thus cannot be optimally represented in Euclidean (or spherical) space. In contrast to the other network layouts plotted here, it does not completely wrap around the sphere due to repulsion from nodes at the other side of the grid. \textbf{(B)} The complexity of connections of networks like "dw256A" may lead the simulation to a local minimum. Most satisfying results were achieved with a high number of simulation steps (here: 500). \textbf{(C)} and \textbf{(D)} Small-world networks generated by the Watt-Strogatz model. In (D) the long chains of nodes fold onto the sphere surface, i.e. its embedding is most probably too low-dimensional. \textbf{(E)} Resulting quality (i.e. uniform node distribution, uniform edge length) is comparable to graph layout algorithms in 2D Euclidean space \cite{hu2015visualizing}. \textbf{(F)} Larger small-world network layouts may contain many edge crossings.  }
\label{examplesSmall}
\end{figure}

\begin{figure}
\begin{adjustwidth}{-6.0cm}{-0mm}
\includegraphics[width=18.0cm]{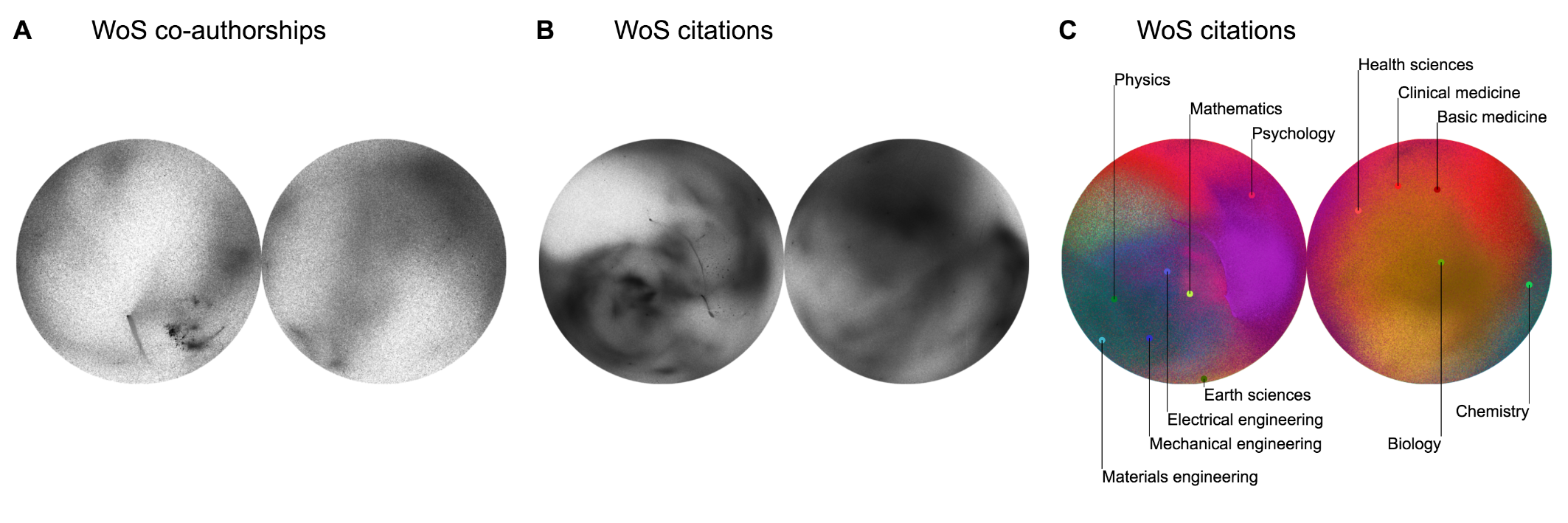}
\end{adjustwidth}
\caption[Spherical layout of large networks]{\textbf{Spherical layout of large networks.}  Network data as described in Table \ref{focaldatasourcestable}. \textbf{(A)} and \textbf{(B)} Spherical coordinates are divided into two half-spheres and projected using a Lambert azimuthal equal-area mapping. Due to the high number of nodes (2.5M for (A) and 38.9M for (B)), positions are visualized as a density plot, where darker areas mean a higher concentration of nodes. \textbf{(C)} Nodes (publications) are colored according to the OECD science \& technology classification, which consists of 42 disciplines. For visual clarity, only the 12 disciplines with the highest number of publications are labeled. }
\label{examplesLarge}
\end{figure}

\begin{figure}
\begin{adjustwidth}{0.0cm}{-0mm}
\includegraphics[width=12.0cm]{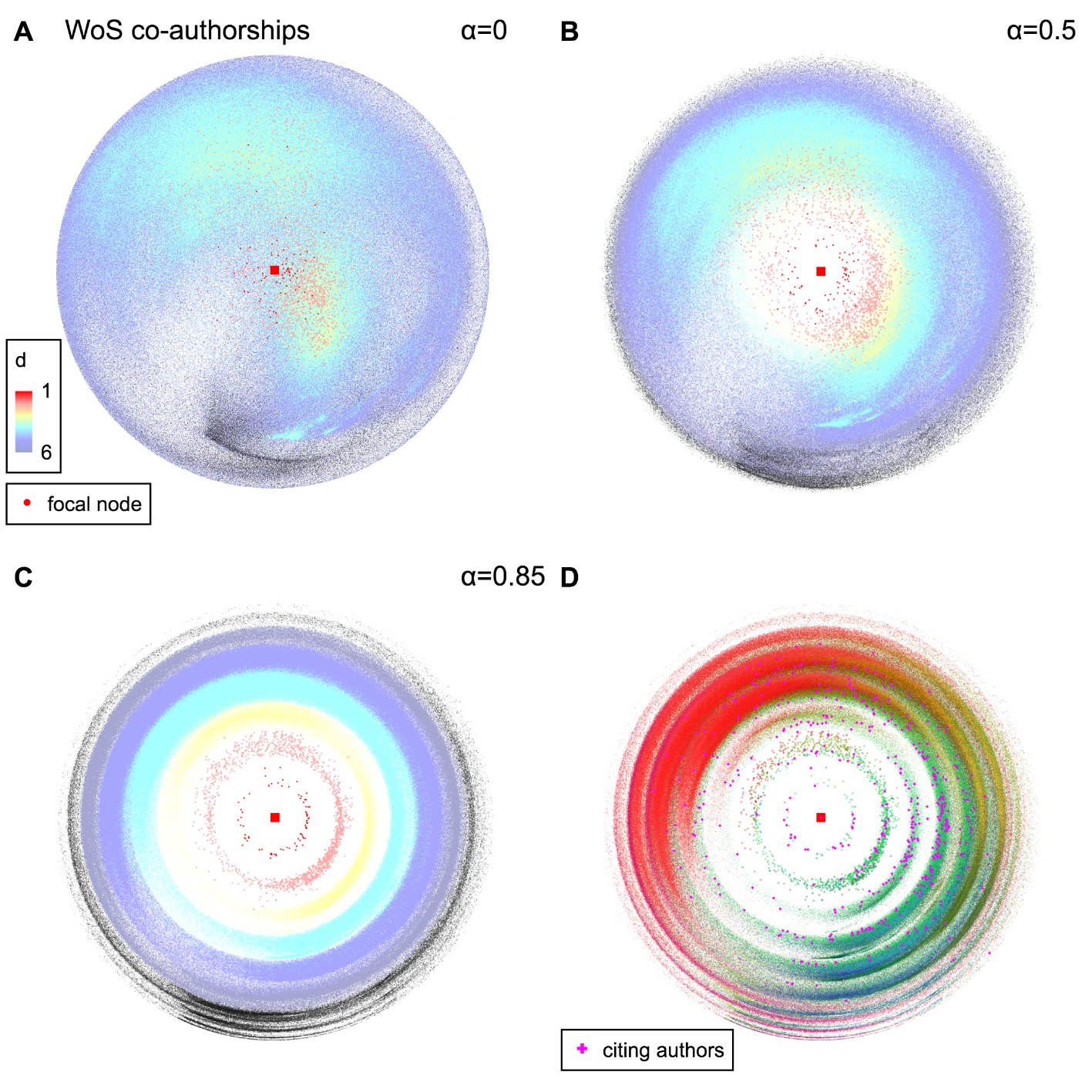}
\end{adjustwidth}
\caption[Focal network layout]{\textbf{Focal network layout.}  A random node (author) was chosen and the sphere rotated such that the node is the center of projection. A Lambert azimuthal equal-area mapping projects the whole surface of the sphere. \textbf{(A)} to \textbf{(C)} Nodes are colored by their network distance $d$ to the focal node. Black nodes have a network distance greater than six. While large parts of the network can already be reached from the focal node within six degrees of separation, we observe in (A) that network distance $d$ does not necessarily correspond well to spherical distance $\theta$. In order to better preserve the network distances to the focal node, all nodes are moved towards their corresponding network distance along the arc that connects a node with the focal node. Parameter $\alpha$ controls how well the network distance is respected (with $\alpha=0$: no change to the original layout, and $\alpha=1$: forming a ring for the set of nodes at each network distance. \textbf{(D)} shows an example of the final focal network layout. Authors of articles that cited articles of the focal author are marked in magenta, and additionally, all nodes are classified by scientific disciplines using the color scheme of Fig.~\ref{examplesLarge} C.  }
\label{focalLayout}
\end{figure}

\subsection*{Qualitative evaluation}

Fig.~\ref{examplesSmall} shows network examples using the force-based algorithm in spherical space. Among possible parameters (number of simulation steps, random seed, maximum step size $\theta_{\max}$, a decrease of $\theta_{\max}$ over time, use of a multi-level layout, multi-level network size increase over time), the number of simulation steps has the most substantial effect on the outcome. We set it to a high 500 steps for smaller networks that have less than 1,000 nodes, and 250 steps for more extensive networks. Depending on the network, increasing the number of steps does not improve the quality of the layout anymore. Since small network layouts are computed within a few seconds, a more generous number of steps is chosen to guarantee the best possible layout. Overall, layout results are comparable to other algorithms in 2D Euclidean space. An advantage of the three-dimensional spherical representation is that the user can rotate and zoom into an area of interest, while still being able to see the rest of the network wrapped around the sphere.

 For larger networks, nodes are represented as a density distribution plotted on the surface of the sphere (Fig.~\ref{examplesLarge}). The example networks display areas of strongly varying density. Some visual artifacts are caused by unusual network structures such as a very high-degree node connected to many low-degree nodes. In order to make sense of the plots, additional information is needed (Fig.~\ref{examplesLarge} C).

 Up to now, we only discussed intermediate steps of the algorithm. An example of a final result depicting a focal network layout is shown in Fig.~\ref{focalLayout}. The focal layout may be generated from any node of the network. Circles around the center (the focal node) are formed by nodes of the same network distance to the focal node. The angular position of a node on these circles can be interpreted as similarity or closeness to other nodes. In Fig.~\ref{focalLayout} C we can see that the community structure of the network (represented by color) becomes visible since author nodes of the same discipline appear on a similar angle and distance to the focal author.

\begin{figure}
\begin{adjustwidth}{0.0cm}{-0mm}
\includegraphics[width=12.0cm]{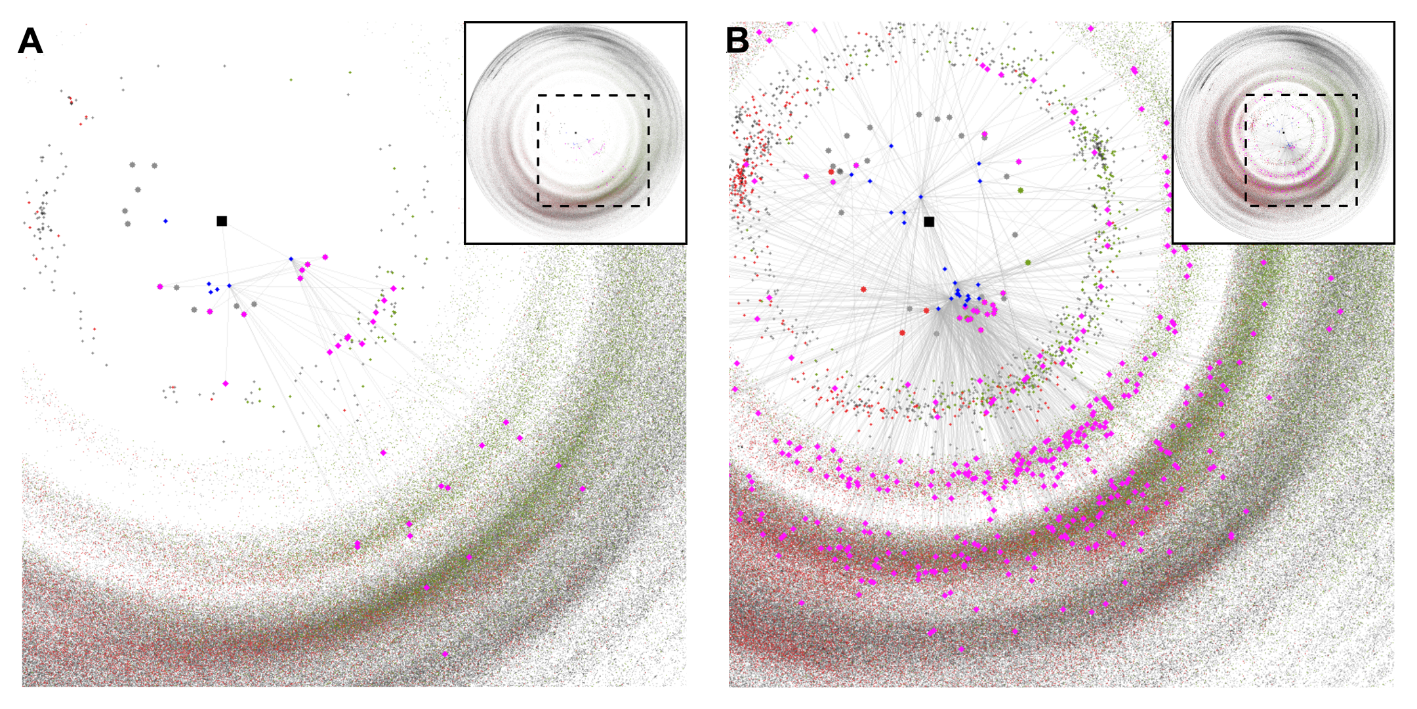}
\end{adjustwidth}
\caption[Dynamic focal network layout]{\textbf{Dynamic focal network layout.}  Example of an evolving collaboration network around a focal author (black node). Citing authors are marked in purple, with a line towards the cited publication (blue). Coordinates of the the focal author's publications are the mean coordinates of all co-authors.}
\label{focalLayoutAnim}
\end{figure}

 For animating an evolving network (Fig.~\ref{focalLayoutAnim}), nodes can be added in a chronological order similar to the described multi-level method. Since the focal node potentially moves between subsequent time steps, angular coordinates are likely not well aligned after the focal node rotates to the projection center. This rotational offset around the projection axis can be corrected by finding the angle that minimizes the coordinate deviations of all nodes in subsequent animation frames.  

\begin{figure}
\begin{adjustwidth}{-6.0cm}{-0mm}
\includegraphics[width=18.0cm]{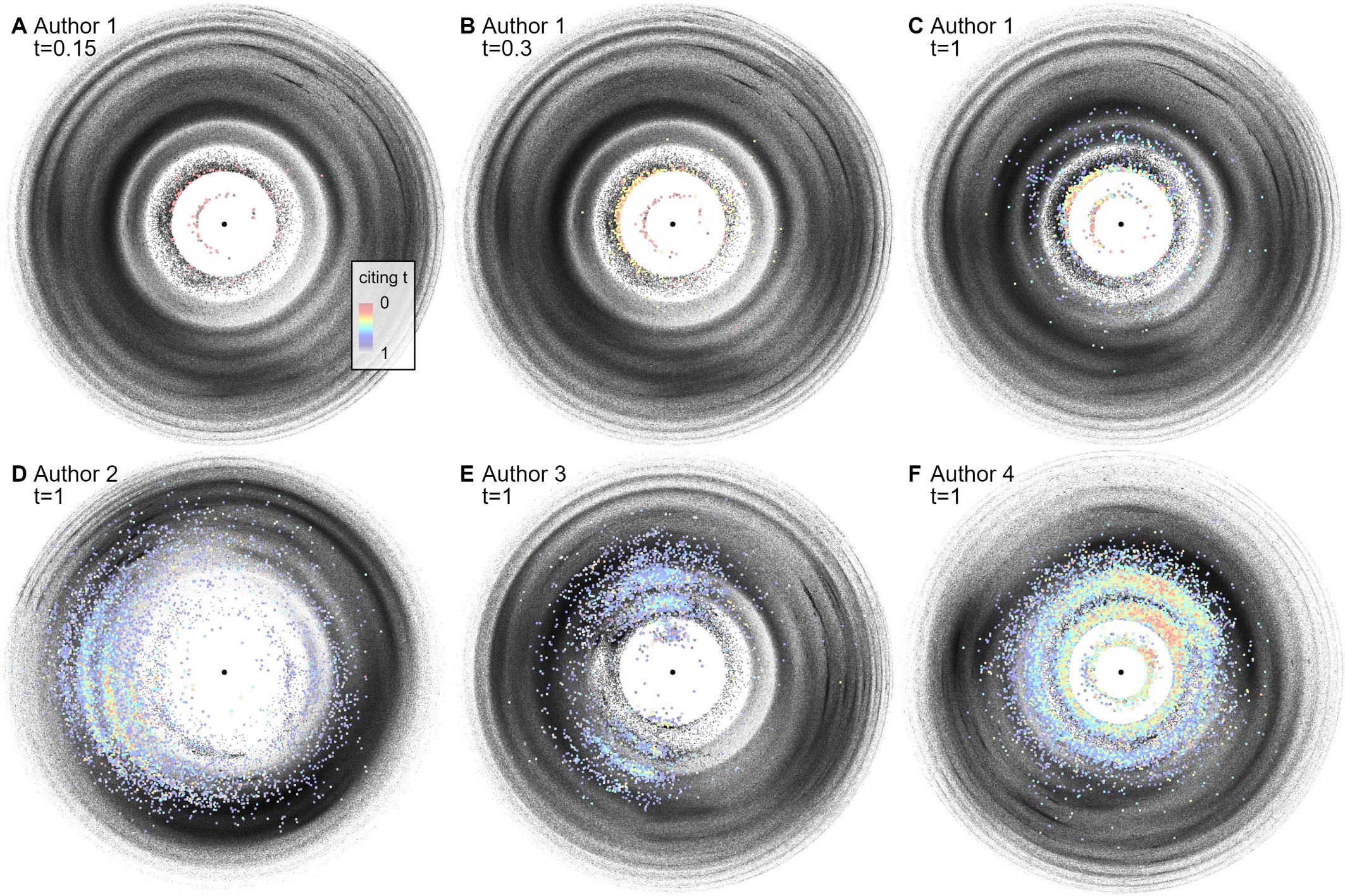}
\end{adjustwidth}
\caption[Visualizing spreading on a focal network layout]{\textbf{Visualizing spreading on a focal network layout.}  Collaboration network around selected authors (MAG data, see Table \ref{focaldatasourcestable}), with authors citing the focal author marked in color. First time of a citation is mapped to a cooling color scheme (start of the focal's author career to latest available year: $t=[0,1] \mapsto$ [red, yellow, green, blue, white]). \textbf{(A-C)} First authors who cite the focal author are mainly at distance 1 or 2. \textbf{(D)} An author with a sparser collaboration network, yet exceptional citation success. Early success concentrated in a smaller community before the author's publications became known to a broader audience. \textbf{(E)} Here, the focal author gained very good visibility in two local communities. \textbf{(F)} An author who is well-connected and also very successful throughout the chosen field of study.}
\label{focalSpreadingExamples}
\end{figure}

 Visualizing the actual spreading phenomena needs additional data, typically containing at which point in time a node became infected or adopted a behavior or an idea and so forth. Here, we explore the role of a scientist's social network (represented by previous collaborations) in the diffusion of ideas by coloring nodes when they cite the focal node for the first time (Fig.~\ref{focalSpreadingExamples}). Indeed, we can find examples where there is a visible spread in the space we have chosen to embed the network. We observe that co-authors and their co-authors are often the first to cite the focal author's work and only later in the career, citations are happening at greater network distance. The spread is not necessarily only away from the cited author (increasing radius), but also along the angular coordinate, indicating a local diffusion along the community structure. Since the co-author network is also evolving, we would expect feedback effects between collaboration activity and citation success. From the visualizations alone, it may be difficult to derive any conclusions on a statistical level, but they may help to create first hypotheses about the spreading process of interest.  

\subsection*{Quantitative evaluation}

\begin{figure}
\begin{adjustwidth}{-6.0cm}{-0mm}
\includegraphics[width=18.0cm]{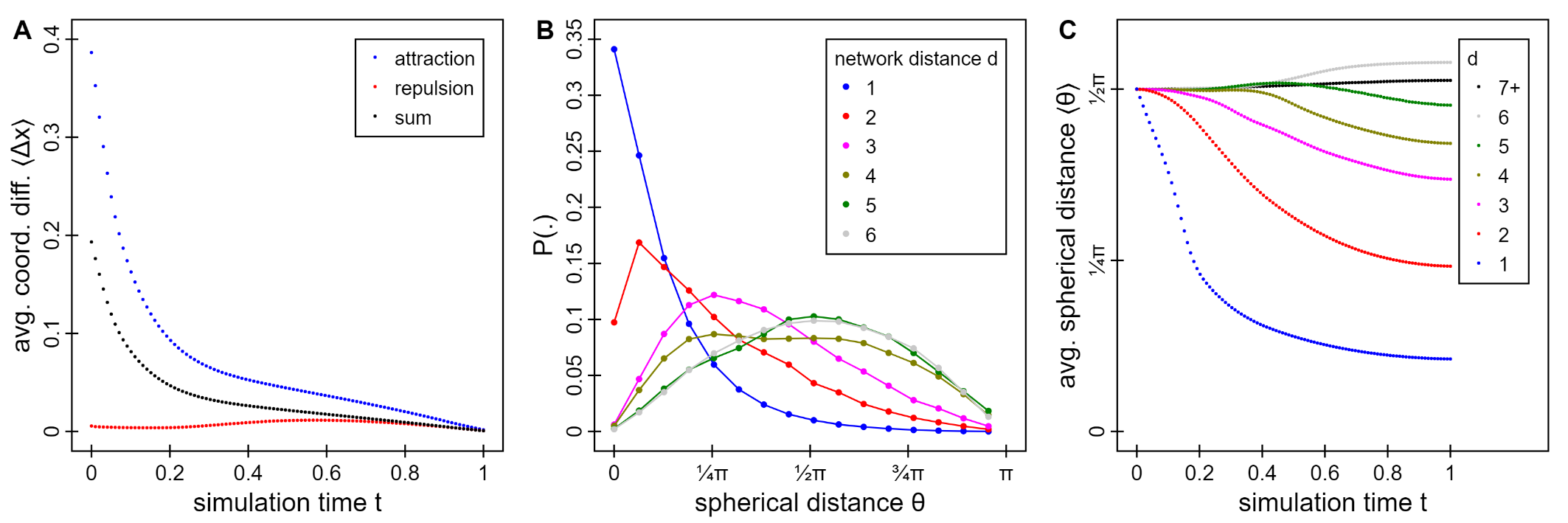}
\end{adjustwidth}
\caption[Error minimization]{\textbf{Error minimization.}  All plots are created using the "WoS co-authorships" network as described in Table \ref{focaldatasourcestable}. \textbf{(A)} Mean coordinate changes per node and simulation step. Attractive forces are the highest at the beginning of the simulation since neighboring nodes are furthest apart. A random position initialization leads to repulsion forces from all directions that cancel out each other and only later, when structure emerges, start to raise, on average. Over time, repulsion and attraction balance out, i.e. pointing in opposite directions, and positions stabilize. \textbf{(B)} Distribution of spherical distances $\theta$ between node pairs grouped by actual network distance $d$, after the last step of the simulation. An optimal layout would generate non-overlapping distributions.  \textbf{(C)} Mean spherical distances during simulation. The average distance between two random positions on the unit sphere is $\frac{\pi}{2}$ and thus all distance levels start at this average value. Node pairs at network distance 1 shorten their spherical distance the fastest, followed by greater network distances, as an indirect consequence of only optimizing distance 1.  }
\label{validation}
\end{figure}

For a quantitative analysis, we need to understand the dynamics of the layout algorithm. Fig.~\ref{validation} A plots the influence of the two force types throughout the layout simulation. Attractive forces bring neighboring nodes closer together, while repulsive forces keep the nodes spread out around the sphere. This means we are only minimizing the distances of nodes pairs that are directly connected by an edge. And indeed, the average spherical distance of node pairs with a network distance of 1 significantly decreases over time, as shown in Fig.~\ref{validation} C. We also observe that node pairs with a higher network distance shorten their spherical distance, which is an indirect consequence of only optimizing directly connected node pairs. The simulation ideally ends where the spherical distance of any node pair corresponds to its network distance (scaled to cover the range of spherical distances from 0 to $\pi$).

\begin{figure}
\begin{adjustwidth}{0.0cm}{-0mm}
\includegraphics[width=12.0cm]{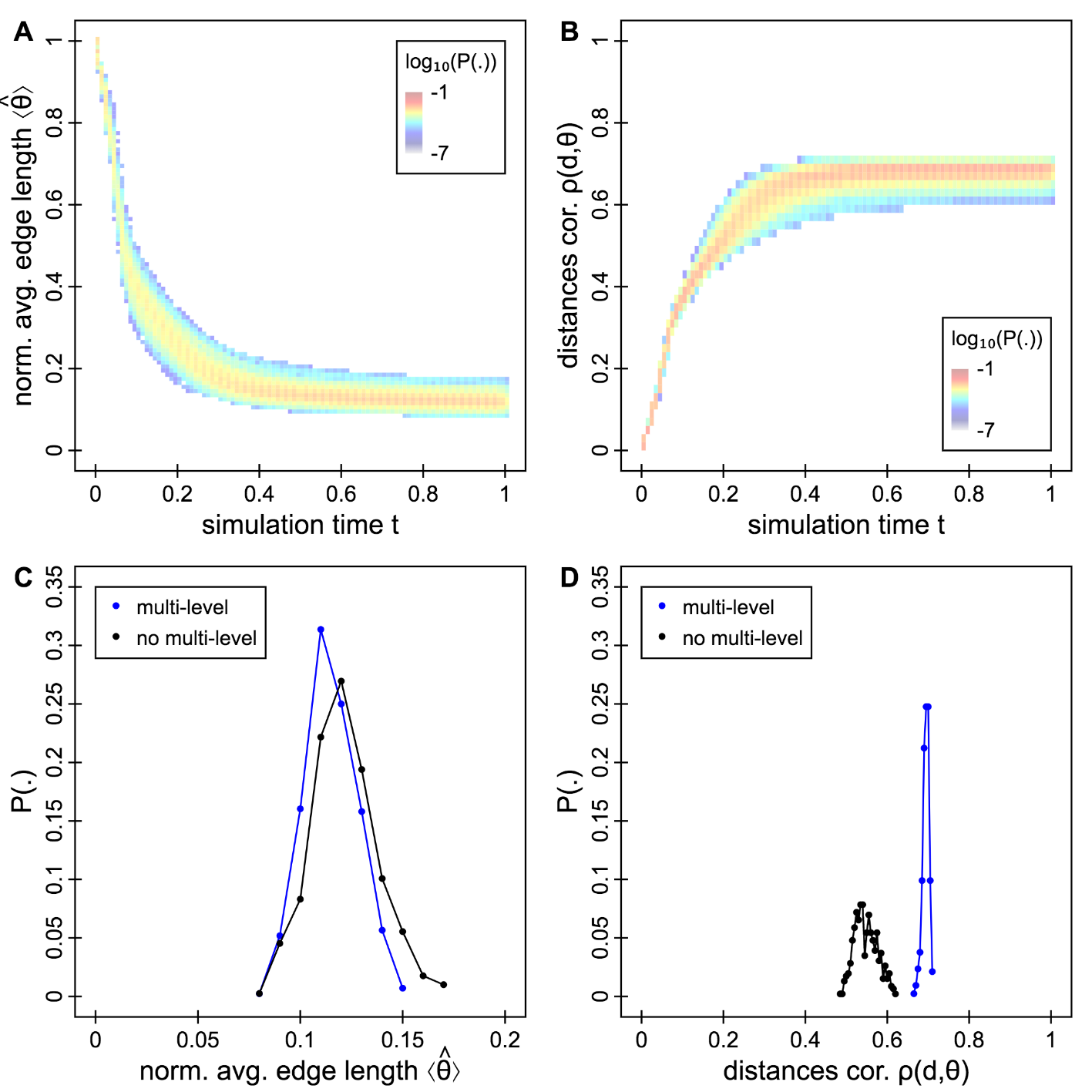}
\end{adjustwidth}
\caption[Variability of different simulation runs and effect of multi-level graph coarsening]{\textbf{Variability of different simulation runs and effect of multi-level graph coarsening.}  Different random start coordinates lead to different outcomes such that the simulation may end up in a local minimum. \textbf{(A)} and \textbf{(B)} are density plots for the two layout quality measures of 400 simulations of "WoS co-authorships" with different random seeds. \textbf{(C)} and \textbf{(D)} Using a multi-level graph coarsening can lead to a significant quality improvement. While the normalized average edge length is varying more severely with different simulation runs than with or without a multi-level approach, correlation of distances highly benefits from using multi-level graphs, which also decreases variability for different seeds.}
\label{localOptimum}
\end{figure}

 A layout quality measure described in \cite{noack2007unified} computes the ratio of average edge length and average distance between all pair of nodes:

\begin{equation}
\langle \hat \theta \rangle = (\sum_i^V \sum_j^{N(i)} \theta_{ij} / \sum_i^V |N(i)|) / (\sum_i^V \sum_{j \neq i}^V \theta_{ij} / |V^{(2)}|)
\end{equation}
Lower values are better. Since it only considers network distance 1, we define another quality measure as the Pearson correlation coefficient between spherical distance $\theta$ and network distance $d$ of any node pair:

\begin{equation}
\rho(d,\theta)= \frac{\operatorname{cov}(d,\theta)}{\sigma(d) \sigma(\theta)}
\end{equation}
For large networks, we need to sample from all possible node pairs. An equal number of samples is drawn from each network distance, otherwise the most important lower distances 1, 2 or 3 would be significantly underrepresented. We also only include distances up to 6, since larger distances (for the type of networks we use here) cannot be expected to deviate much from a random distance distribution. Results for individual networks are shown in Table \ref{focaldatasourcestable}. Fig.~\ref{localOptimum} shows the variance of the quality measure when starting with different random seeds. For large networks, using a multi-level refinement of the network during simulation results in a significant improvement of $\rho(d,\theta)$. In contrast, a low $\langle \hat \theta \rangle$ is easier to achieve than a high $\rho(d,\theta)$ since the former is looking only at local structure (the immediate neighborhood), while the latter includes higher distances, too. Thus, Fig.~\ref{localOptimum} D suggests that a multi-level approach is better at avoiding local minima.

\begin{figure}
\begin{adjustwidth}{6.0cm}{-0mm}
\includegraphics[width=6.0cm]{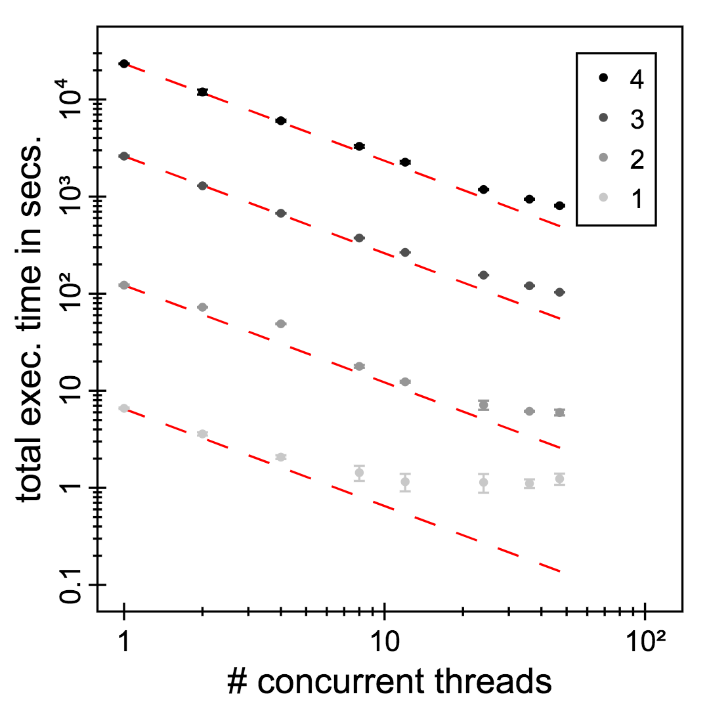}
\end{adjustwidth}
\caption[Scalability]{\textbf{Scalability.}   Execution duration for network layouts of different sizes and degree of parallel computing. 1. WoS Sociology, 2. WoS Math, 3. WoS Biology, 4. all WoS co-authorships, measuring the total time to execute 100 simulation steps. Shown is the average time out of 3 simulation runs, error bar is standard deviation. The red line marks an ideal speed-up, i.e. a doubling in computing resources would mean half the execution time. For small networks, increasing the number of threads has diminishing returns and does not go below a second as the total execution time. Larger networks can benefit from multi-core systems and show an excellent speed-up almost at the theoretical maximum. Tested on a 2x12 core machine with a maximum of 48 threads.  }
\label{scalability}
\end{figure}

\subsection*{Scalability}

Since the algorithm offers great potential for parallelization, the implementation in Java is optimized for running on multi-core and multi-machine hardware. The triangle quadtree data structure is only initialized once, and the mass center of each triangle is reset before each simulation step. Adding a node to the tree is an independent operation and can be parallelized, only the concurrent modification of the mass center and the list of individual nodes at the tree leaves need to be synchronized. Position updates are also computed independently. Potentially already updated positions cannot be used within the same step since the tree structure would need to be re-initialized. Updated positions may also not be available immediately when multiple machines are used. This is a compromise for the sake of scalability that may result in nodes jumping forward and backward or the whole network drifting in one direction. For a multi-machine setup, quadtree initialization is done on a single multi-core machine. Then, the computation of position updates is distributed to multiple machines. The biggest challenge for maintaining good scalability is efficient data synchronization over the network between individual machines. On each machine, all data has to be available. The complete network (neighborhood of each node), used by the attraction force computation, is distributed once and does not change during a simulation step. The quadtree data structure only needs to update mass centers and lists of individual nodes at the leaves. All updated node positions need to be available on all machines.

 Results for a multi-core setup (Fig.~\ref{scalability}) show excellent scalability, especially for large networks. On multi-machine hardware, it is much more difficult to achieve good scalability due to communication overhead and load balancing. For the WoS co-authorship network, we measure a run-time of 364 seconds on a cluster with ten machines, which is 5.3 times faster than on a single machine. For the largest network data, the citation network, the run-time is only decreased by a factor of 4.3 on a cluster with 20 machines.

 Once a global network layout has been established, focal network layouts can be computed efficiently. Selecting a node on the MAG co-authorship network and drawing a focal network as depicted in Fig.~\ref{focalSpreadingExamples} typically takes less than a second. Implemented as a web server, requests return a raster image of the focal network. Most time is spent on the single-source distances computation, image compression and data transfer to the client. Additionally, coordinates of selectable nodes need to be transferred to the browser. In our example, we only allow to select co-authors and citing authors, which considerably reduces the amount of data sent to the client. Any other node may be retrieved by a separate text search index that is created from further node metadata. 

\section*{Conclusion}

This paper proposes an algorithm that computes an interactive layout for large-scale network data from the perspective of a single node. The resulting visualization puts a focal node in the center of a polar coordinate system, where the radial coordinate of a node corresponds to the network distance to the focal node, and differences in the angular coordinate correspond to network distances between any nodes. Thus, it is precisely suited for gaining insight into phenomena that can be characterized by network distance. Especially for large networks with a low average shortest path, a global network layout would mostly only give an idea of an approximate community structure, where a focal network layout can reveal more about the position of a node in the network and its immediate neighborhood.

 The proposed algorithm only finds optimal node coordinates and does not prescribe how to integrate additional information on the plot. The examples presented here encode data into node coloring, which may not be ideal for all applications. An extension to allow several nodes to be in focus would cover more possible scenarios. It is also not yet investigated how weighted networks would change the visualization.   In case of billions of nodes, an improved multi-machine implementation would become necessary.

\bibliography{literature}

\end{document}